\documentclass[aps,prl,twocolumn,superscriptaddress,reprint]{revtex4-2}
\usepackage{graphicx}
\usepackage{amssymb, amsmath}
\usepackage{booktabs}
\usepackage[usenames]{color}
\usepackage{bm}
\usepackage{multirow}
\usepackage{dcolumn}
\usepackage{hyperref}
\usepackage{enumerate}
\hypersetup{
    colorlinks=true,
    linkcolor=blue,
    filecolor=gray,      
    urlcolor=blue,
    citecolor=blue,
}

\begin{document}

\title{Deep-Learning Density Functional Theory Hamiltonian for Efficient {\it ab initio} Electronic-Structure Calculation}

\affiliation{State Key Laboratory of Low Dimensional Quantum Physics and Department of Physics, Tsinghua University, Beijing, 100084, China}
\affiliation{Institute for Advanced Study, Tsinghua University, Beijing 100084, China}
\affiliation{School of Physics, Peking University, Beijing 100871, China}
\affiliation{Frontier Science Center for Quantum Information, Beijing, China}
\affiliation{Beijing Academy of Quantum Information Sciences, Beijing 100193, China}
\affiliation{RIKEN Center for Emergent Matter Science (CEMS), Wako, Saitama 351-0198, Japan}
\affiliation{These authors contributed equally}

\author{He \surname{Li}}
\affiliation{State Key Laboratory of Low Dimensional Quantum Physics and Department of Physics, Tsinghua University, Beijing, 100084, China}
\affiliation{Institute for Advanced Study, Tsinghua University, Beijing 100084, China}
\affiliation{These authors contributed equally}

\author{Zun \surname{Wang}}
\affiliation{State Key Laboratory of Low Dimensional Quantum Physics and Department of Physics, Tsinghua University, Beijing, 100084, China}
\affiliation{These authors contributed equally}

\author{Nianlong \surname{Zou}}
\affiliation{State Key Laboratory of Low Dimensional Quantum Physics and Department of Physics, Tsinghua University, Beijing, 100084, China}
\affiliation{These authors contributed equally}

\author{Meng \surname{Ye}}
\affiliation{State Key Laboratory of Low Dimensional Quantum Physics and Department of Physics, Tsinghua University, Beijing, 100084, China}

\author{Runzhang \surname{Xu}}
\affiliation{State Key Laboratory of Low Dimensional Quantum Physics and Department of Physics, Tsinghua University, Beijing, 100084, China}

\author{Xiaoxun \surname{Gong}}
\affiliation{State Key Laboratory of Low Dimensional Quantum Physics and Department of Physics, Tsinghua University, Beijing, 100084, China}
\affiliation{School of Physics, Peking University, Beijing 100871, China}

\author{Wenhui \surname{Duan}}
\email{duanw@tsinghua.edu.cn}
\affiliation{State Key Laboratory of Low Dimensional Quantum Physics and Department of Physics, Tsinghua University, Beijing, 100084, China}
\affiliation{Institute for Advanced Study, Tsinghua University, Beijing 100084, China}
\affiliation{Frontier Science Center for Quantum Information, Beijing, China}
\affiliation{Beijing Academy of Quantum Information Sciences, Beijing 100193, China}

\author{Yong \surname{Xu}}
\email{yongxu@mail.tsinghua.edu.cn}
\affiliation{State Key Laboratory of Low Dimensional Quantum Physics and Department of Physics, Tsinghua University, Beijing, 100084, China}
\affiliation{Frontier Science Center for Quantum Information, Beijing, China}
\affiliation{RIKEN Center for Emergent Matter Science (CEMS), Wako, Saitama 351-0198, Japan}

\begin{abstract}
The marriage of density functional theory (DFT) and deep learning methods has the potential to revolutionize modern computational materials science. Here we develop a deep neural network approach to represent DFT Hamiltonian (DeepH) of crystalline materials, aiming to bypass the computationally demanding self-consistent field iterations of DFT and substantially improve the efficiency of {\it ab initio} electronic-structure calculations. A general framework is proposed to deal with the large dimensionality and gauge (or rotation) covariance of DFT Hamiltonian matrix by virtue of locality and is realized by the message passing neural network for deep learning. High accuracy, high efficiency and good transferability of the DeepH method are generally demonstrated for various kinds of material systems and physical properties. The method provides a solution to the accuracy-efficiency dilemma of DFT and opens opportunities to explore large-scale material systems, as evidenced by a promising application to study twisted van der Waals materials.
\end{abstract}

\maketitle
\section{Introduction}

Nowadays {\it ab initio} calculations based on density functional theory (DFT)~\cite{Hohenberg1964,Kohn1965} have become indispensable to scientific research in physics, materials science, chemistry and biology~\cite{Jones2015}, and meanwhile deep learning based on neural network has revolutionized many disciplines ranging from computer vision, natural language processing to scientific discoveries~\cite{LeCun2015, Jordan2015, Carleo2019}. The marriage of these two paramount fields creates an emerging direction of deep-learning {\it ab initio} calculation~\cite{Behler2007, Schutt2018, Zhang2018, Klicpera2020, Oliver2021, Brockherde2017, Grisafi2019, Chandrasekaran2019, Tsubaki2020, Grisafi2018, Gu2020, Schutt2019, Unke2021, Nagai2020, Dick2020, Kirkpatrick2021, Mills2019, Zubatiuk2021}, which sheds light on the development of computational materials science. One critical problem of DFT is that it is computationally rather demanding and hardly applicable to material systems with over thousands of atoms for routine calculations. One may employ more efficient algorithms (e.g., the linear-scaling methods~\cite{Goedecker1999}), but usually at the expense of decreasing accuracy and transferability. In principle, benefiting from the remarkable expressive power, deep neural network can learn well from DFT results and be applied to bypass computationally expensive steps. The accuracy-efficiency dilemma of DFT thus might be solved via deep learning, which facilitates the exploration of various important systems, including defects, disorders, interfaces, heterostructures, quasi-crystals, twisted van der Waals (vdW) materials, etc.

Until now tremendous efforts have been devoted to developing deep-learning potential that learns the inter-atomic interaction or the potential energy from DFT by neural network~\cite{Behler2007, Schutt2018, Zhang2018, Klicpera2020, Oliver2021}. By applying this potential, molecular dynamics (MD) simulations can exhibit the efficiency of classical MD with the {\it ab initio} accuracy. Thus the research scope of material simulation is largely expanded. Naturally one would like to generalize the deep-learning idea from the atomic simulation level to the electronic structure simulation level. The most fundamental quantity to be learned is the DFT Hamiltonian~\cite{Hegde2017}, from which almost all electron-related physical quantities in the single-particle picture can be derived, such as charge density, band structure, Berry phase, and physical responses to electromagnetic fields. Instead of studying these physical quantities separately~\cite{Brockherde2017, Grisafi2019, Chandrasekaran2019, Tsubaki2020, Grisafi2018, Gu2020}, applying the deep learning method to DFT Hamiltonian is a more essential and challenging task. In contrast to gauge-invariant quantities, the DFT Hamiltonian matrix transforms covariantly (i.e. equivariantly) under changes of coordinate, basis and gauge, thus demanding the design of a gauge (or rotation) covariant neural network~\cite{Nathaniel2018, Anderson2019, Fuchs2020}. Moreover, when applying a neural network to represent the relation between material structure and DFT Hamiltonian for large-scale material structures, the number of independent variables and the dimension of Hamiltonian matrix would both become infinitely large. Previous works designed neural networks to study DFT Hamiltonian of small molecules~\cite{Schutt2019, Unke2021}. Another work considered a specific one-dimensional material and circumvented the gauge issue by learning energy eigenvalues~\cite{Gu2020}. Despite these preliminary attempts, developing deep learning DFT Hamiltonian for electronic structure calculation of large-scale material systems remains elusive.

In this work, we propose a general theoretical framework of deep learning  DFT Hamiltonian (DeepH) to study crystalline materials via the message passing neural network. The challenging issues related to the (infinitely) large dimensionality and gauge (or rotation) covariance of DFT Hamiltonian matrix are solved by virtue of locality, including the use of local coordinate, local basis transformation, and localized orbitals as basis functions. We systematically tested the capability of the DeepH method by studying various representative materials with flat or curved structures, formed by strong chemical bonds or weak vdW bonds, containing single or multiple elements, excluding or including spin-orbit coupling (SOC), etc. The example studies consistently demonstrated the high accuracy of DeepH not only in the construction of DFT Hamiltonian (minor error $\sim$meV) but also in calculations of band- and wavefunction-related physical quantities. Remarkably, the DeepH method performs extremely well in investigating twisted vdW materials in terms of accuracy, transferability and efficiency, which is an advantage to build twisted materials database. Our method is expected to be universal, applicable to periodic or non-periodic systems, and could find useful applications in computational materials science.

\section{Results}

\subsection{Theoretical framework of DeepH}
\begin{figure*}[h!]
\includegraphics[width=0.75\linewidth]{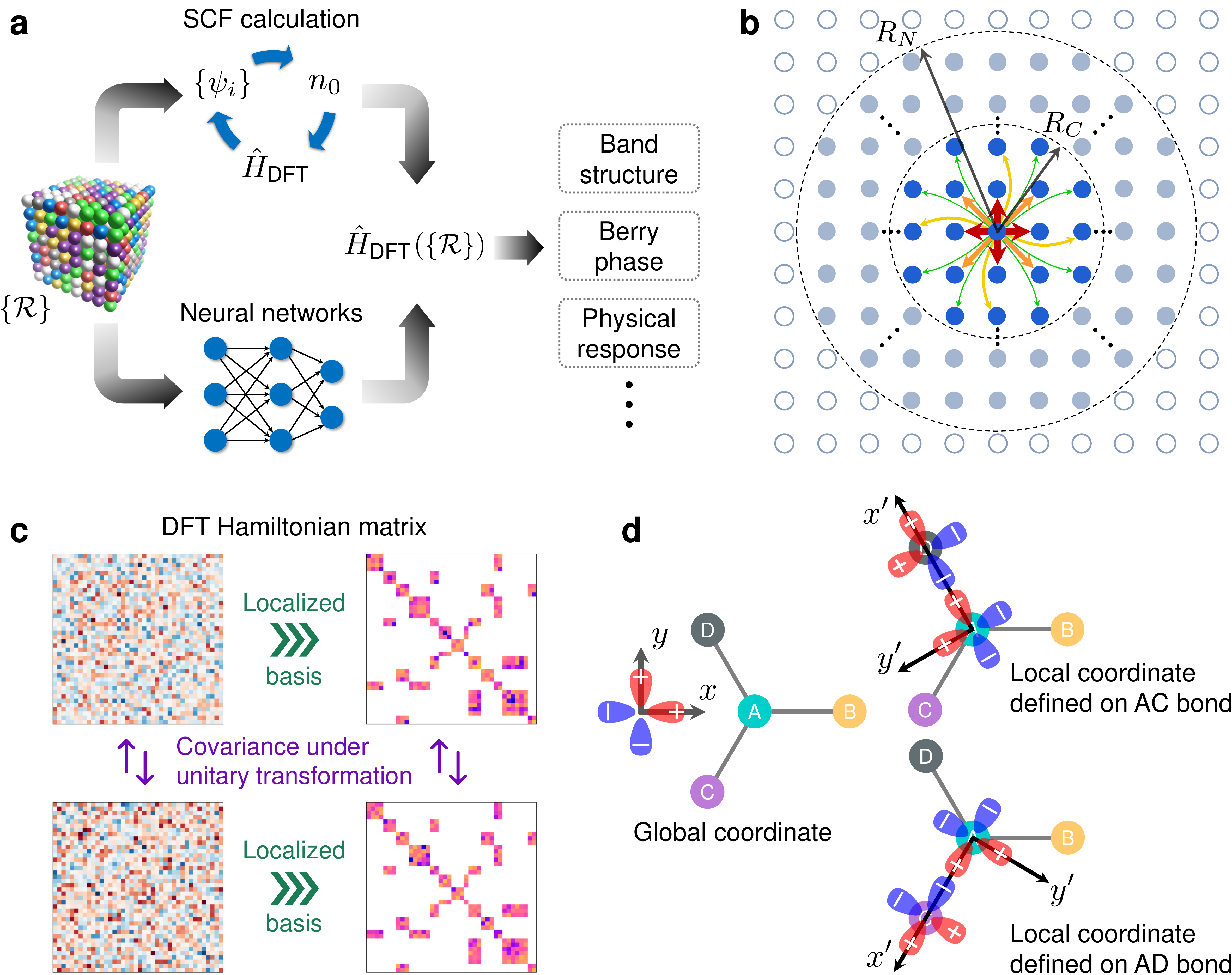}
\caption{Learning DFT Hamiltonian $\hat{H}_{\text{DFT}}$ by virtue of locality. \textbf{a} $\hat{H}_{\text{DFT}}$ as a function of material structure (i.e., atomic coordinate $\{\mathcal{R}\}$), which can be obtained by self-consistent field (SCF) calculation or alternatively learned by neural network for efficient {\it ab initio} electronic-structure calculation. \textbf{b} Use of nearsightedness principle of electronic matter to learn $\hat{H}_{\text{DFT}}$, whose matrix elements in localized basis are nonzero between neighboring atoms (within $R_C$) and influenced only by neighborhood (within $R_N$). \textbf{c} Schematic diagram showing properties of DFT Hamiltonian matrix, which is generally dense and becomes sparse in localized basis and changes covariantly under unitary transformation. \textbf{d} Illustration of rotation transformation for a four-atom structure with $p_{x,y}$ orbitals in varying coordinates.}
\label{fig1}
\end{figure*}

One of the most fundamental problems in quantum physics is to solve the Schr\"{o}dinger equation for interacting electrons of matter, so as to predict material properties from first principles. DFT~\cite{Hohenberg1964,Kohn1965} has been highly recognized for the purpose, which replaces the complicated many-body problem by a simpler auxiliary one $\hat H_{\text{DFT}} |\psi\rangle = \mathcal{E}|\psi\rangle$ describing non-interacting electrons with interacting density~\cite{Martin2004}, where $\hat H_{\text{DFT}}$ is the DFT Hamiltonian operator, $\mathcal{E}$ and $|\psi\rangle$ are the Kohn-Sham eigenvalue and eigenstate, respectively. Typically {\it ab initio} DFT Hamiltonian $\hat H_{\text{DFT}}$ is obtained via self-consistent field calculations, followed by calculations of material properties (Fig. \ref{fig1}a). According to the Hohenberg-Kohn theorem~\cite{Hohenberg1964}, there is a one-to-one correspondence between the external field determined by the material structure $\{\mathcal{R}\}$ and $\hat H_{\text{DFT}}$, implying a mapping function: $\{\mathcal{R}\}\mapsto \hat{H}_{\text{DFT}}(\{\mathcal{R}\})$. The generic form of $\hat{H}_{\text{DFT}}(\{\mathcal{R}\})$, however, is too complicated to be expressed analytically, but can be represented by the DeepH method. For generally non-periodic crystalline materials containing an infinite number of atoms, $\hat{H}_{\text{DFT}}(\{\mathcal{R}\})$ has an infinite number of independent variables in $\{\mathcal{R}\}$. Therefore, the DFT Hamiltonian matrix may have an infinitely large dimension and the matrix is invariant under atom permutation and translation, and covariant under rotation and gauge transformations (Fig. \ref{fig1}b). In this sense, learning the DFT Hamiltonian is much more challenging than learning scalar physical quantities, such as total energy~\cite{Behler2007, Schutt2018, Zhang2018, Klicpera2020}.

Next, we show that the problem of learning DFT Hamiltonian, though looking formidable, can be solved by virtual of the locality.  As revealed by W. Kohn and others, local physical properties do not respond to distant changes of external potential due to the destructive interference between many-particle eigenstates~\cite{Kohn1996,Prodan2005}. This implies a widely applicable principle of locality or ``nearsightedness'' in electronic matter. Thus, there is no need to study the entire system at once, and only information of neighborhood is relevant for learning DFT Hamiltonian (Fig. \ref{fig1}b).

A proper selection of basis sets is essential to DeepH. DFT calculations usually use plane waves or localized orbitals as basis functions. The latter is compatible with the locality and non-periodicity nature of the problem and thus will be employed. Then $\hat{H}_{\text{DFT}}$ is expressed as a sparse matrix (Fig. \ref{fig1}c) benefiting from the local or semilocal property of the Kohn-Sham potential. The matrix element $H_{i\alpha, j\beta}$ ($\alpha, \beta$ refer to localized orbitals centered at atoms $i,j$) vanishes when the distance between atoms $i$ and $j$ is larger than a cutoff radius $R_C$. $R_C$ is determined by the spread of localized orbitals, which is on the order of angstrom, much smaller than the nearsightedness length $R_N$ (Fig. \ref{fig1}b). We suggest using non-orthogonal atomic-like orbitals. They are typically more localized than orthogonal ones by circumventing the conflicting requirement of localization and orthogonalization~\cite{Wang2019}. Moreover, their gauge is system-independent and the rotation transformation is well described by spherical harmonics. In contrast, the widely used Wannier functions do not possess such kinds of advantages~\cite{Nicola2012}. By taking advantage of the sparseness and nearsightedness, only Hamiltonian matrix blocks $H_{ij}$ between neighboring atoms (within $R_C$) have to be learned, and only information of neighborhood of atoms $i$ and $j$ (within $R_N$) is relevant to learning $H_{ij}$.

A critical issue is to deal with covariant transformations of DFT Hamiltonian matrix. The Hamiltonian matrix itself is not physically observable, which changes covariantly when varying coordinate, basis or gauge globally or locally. Taking a four-atom structure as an example (Fig. \ref{fig1}d), a global rotation of atomic structure (or basis functions) changes the DFT Hamiltonian matrix. The new Hamiltonian matrix is related to the original one by a rotation transformation. The local transformation is less obvious. In this example structure, the atom pairs AB, AC and AD share the same local chemical environment, whose Hamiltonian matrix blocks are related to each other by rotation transformations. Specifically, the transformed Hamiltonian matrix block $H^\prime_{\mathrm{AC}}$ ($H^\prime_{\mathrm{AD}}$) coincides with $H_{\mathrm{AB}}$ under a clockwise rotation of basis functions by 120$^{\circ}$ (240$^{\circ}$) for AC (AD). In infinite crystalline materials, we may encounter atom pairs with varying orientations. Thus it is difficult (if not impossible) to learn the covariant relations by neural network via data augmentation~\cite{Schutt2019}. Here, we propose a strategy to help DeepH work efficiently and accurately via local coordinate (details described in Supplementary Section 3~\cite{supp}), in which the locally transformed Hamiltonian matrix blocks $H_{ij}^\prime$ are invariant under rotation (Fig. \ref{fig1}d). By changing coordinate from local back to global, a rotation (or basis) transformation is applied to $H_{ij}^\prime$. Then the obtained $H_{ij}$ will naturally satisfy the covariant requirement.

\subsection{Neural network architecture of DeepH}

\begin{figure*}[h!]
\includegraphics[width=0.75\linewidth]{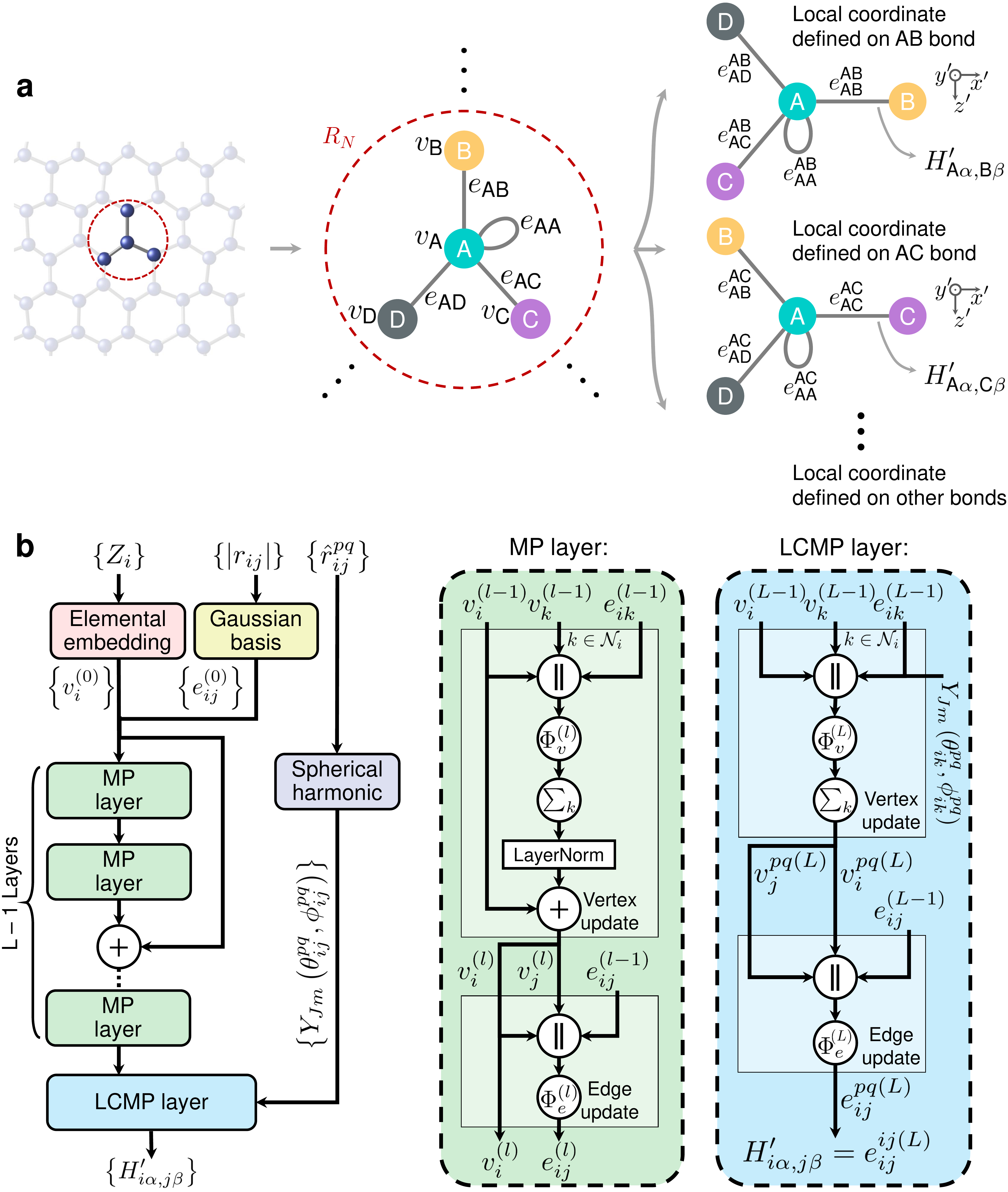}
\caption{Crystal graph and message passing (MP) neural network including $L$ layers employed by DeepH. \textbf{a} Crystal graph with vertices $v_i$ and edges $e_{ij}$ used for MP neural network. For simplicity only edges connected to the nearest neighbors are shown. $R_C$ denotes the cut-off radius. In the local coordinate MP (LCMP) layer, different crystal graphs with new edges $e_{ij}^{pq}$ are applied for different local coordinates defined on varying atom pairs $pq$. \textbf{b} Architecture and workflow of deep neural network, including $L - 1$ MP layers with atomic numbers $\{Z\}$ and interatomic distances $\{|r_{ij}|\}$ as initial inputs and one LCMP layer using additional orientation information $\{\hat{r}_{ik}^{pq}\}$ relative to different local coordinates.}
\label{fig2}
\end{figure*}

Next we present a deep neural network representation of DFT Hamiltonian based on message passing neural network (MPNN)~\cite{Justin2017} that is widely applied for materials studies~\cite{Schutt2018, Schutt2017, Xie2018, Klicpera2020, Wang2021}. Rules of constructing crystal graphs and MPNN are illustrated in Fig. \ref{fig2}a. Each atom is represented by a vertex and atom pairs (with a distance smaller than $R_C$) by edges. MPNN will use edge embeddings to represent $H_{ij}^\prime$. Self-loop edges are added in the graph to consider intra-site couplings. Let $v_{i}$ and $e_{ij}$ denote the vertex feature of atom $i$ and edge feature of atom pair $ij$, respectively. The initial vertex features are the embeddings of atomic number $Z_i$, and the initial edge features are the interatomic distance $\left|r_{ij}\right|$ expanded with Gaussian basis centered at different points $r_n$,
\begin{align}
v^{(0)}_{i} &= \text{Embedding}(Z_i),\\
e^{(0)}_{ij} &= \exp(-(\left|r_{ij}\right| - r_n)^2/\sigma^2).
\end{align}

The architecture and workflow of MPNN are presented in Fig. \ref{fig2}b. In a message passing (MP) layer, the vertex and edge features are updated successively as follows:
\begin{align}
\label{vertex_update}
v_i^{(l)} &= \text{LayerNorm}\left(\sum_{k\in\mathcal{N}_i}\Phi^{(l)}_v\left(z_{ik}^{(l - 1)}\right)\right) + v_i^{(l-1)},\\
\label{edge_update}
e_{ij}^{(l)} &= \Phi^{(l)}_e\left(v_i^{(l)}\parallel v_j^{(l)}\parallel e_{ij}^{(l - 1)}\right),
\end{align}
where $\mathcal{N}_i$ is a set containing neighboring vertices with edge connection to vertex $i$, $\parallel$ denotes the concatenation of feature vectors, the superscript $l$ refers to the $l$-th MP layer, $z_{ik}^{(l - 1)} \equiv v_i^{(l - 1)}\parallel v_k^{(l - 1)}\parallel e_{ik}^{(l - 1)}$ is the concatenation of vertex and edge features of neighborhood, layer normalization ~\cite{Jimmy2016} is employed to improve training efficiency, and $\Phi^{(l)}_v$ and $\Phi^{(l)}_e$ are neural networks for updating vertex and edge features, respectively. The local chemical environment of neighborhood within $R_C$ will be aggregated in an MP layer. As MP layers are stacked, more and more information of the distant chemical environment will be aggregated into the features, enabling the learning of $H_{ij}^\prime(\{\mathcal{R}\}_N)$.

A problem about local coordinate should be remarked. Since the local coordinate is defined for each edge according to its local chemical environment, sometimes minor modifications of local structures could substantially change the coordinate axes, making the transformed $H_{ij}^\prime$ considerably different and thus leading to the inaccuracy of deep learning~\cite{supp}. We find that the problem is solvable by introducing one local coordinate message passing (LCMP) layer after several MP layers. In the LCMP layer, orientation information (unit vector $\hat{r}_{ik}^{pq}$) of bond $ik$ relative to the local coordinate defined for edge $pq$ is added into the initial edge features, where $i$,$k$,$q$,$p$ are all atomic indices. $\theta^{pq}_{ik}$ and $\phi^{pq}_{ik}$ are the corresponding polar and azimuthal angles, respectively, of the reference atoms $i$,$k$,$q$,$p$. The orientation information based on bonds between the central atom and its neighbors was introduced for the study of total energy by Zhang {\it{et al}}.~\cite{Zhang2018}. The vertex and edge features ($v_{i}^{pq(L)}$ and $e^{pq(L)}_{ij}$) defined for local coordinate $pq$ are updated as follows: 
\begin{align}
v_{i}^{pq(L)} &= \sum_{k\in\mathcal{N}_i}\Phi^{(L)}_v\left(z_{ik}^{(L - 1)} \parallel\left\{Y_{Jm}(\theta^{pq}_{ik}, \phi^{pq}_{ik})\right\}\right),\\
e^{pq(L)}_{ij} &= \Phi^{(L)}_e\left(v^{pq(L)}_{i}\parallel v^{pq(L)}_{j}\parallel e_{ij}^{(L - 1)}\right),
\end{align}
where a set of real spherical harmonic functions $\{Y_{Jm}\}$ are used to capture orientation information, and $e^{ij(L)}_{ij}$ will be used to represent $H_{ij}^\prime$. Note that the introduction of LCMP layer into DeepH is critical to improving prediction accuracy according to our test (Supplementary Table 3). Finally $H_{ij}$ is calculated from $H_{ij}^\prime$ via rotation transformation. The neural network of DeepH is trained by DFT data and then applied to predict the DFT Hamiltonian for unseen atomic structures, which can bypass the time-consuming DFT self-consistent calculation and enable efficient electronic-structure calculations.

\subsection{Capability of DeepH}

Example studies are performed on various representative materials, including graphene, MoS$_2$ and their curved counterparts (i.e. nanotubes),
as well as Moir\'e-twisted materials with negligible or strong SOC. The DFT Hamiltonian is computed by using 13, 19, 13, and 19 non-orthogonal atomic-like basis functions for C, Mo, S, and Bi, respectively. The MPNN model including five MP layers followed by one LCMP layer is trained by minimizing the loss function defined as the mean squared errors of $H_{i\alpha, j\beta}^\prime$. Once $\hat H_{\text{DFT}} (\{\mathcal{R}\})$ is learned by the neural network, various kinds of physical properties, such as band structure, Berry phase, and physical responses to external fields, can be predicted while bypassing DFT self-consistent calculations (Fig. \ref{fig1}a). To check the reliability of our method, we study eigen-energy-based quantities (density of states (DOS) or bands) as well as wavefunction-related properties (optical transition and shift current). Shift current is of particular interest as an important photovoltaic effect generated by nonlinear optical progress and closely related to topological quantities (e.g., Berry phase and curvature)~\cite{Morimoto2016, Wang2017, Wang2019}. The linear and nonlinear optical responses are studied using the methods we developed~\cite{Wang2017, Wang2019}.

\begin{figure}[h!]
\includegraphics[width=\linewidth]{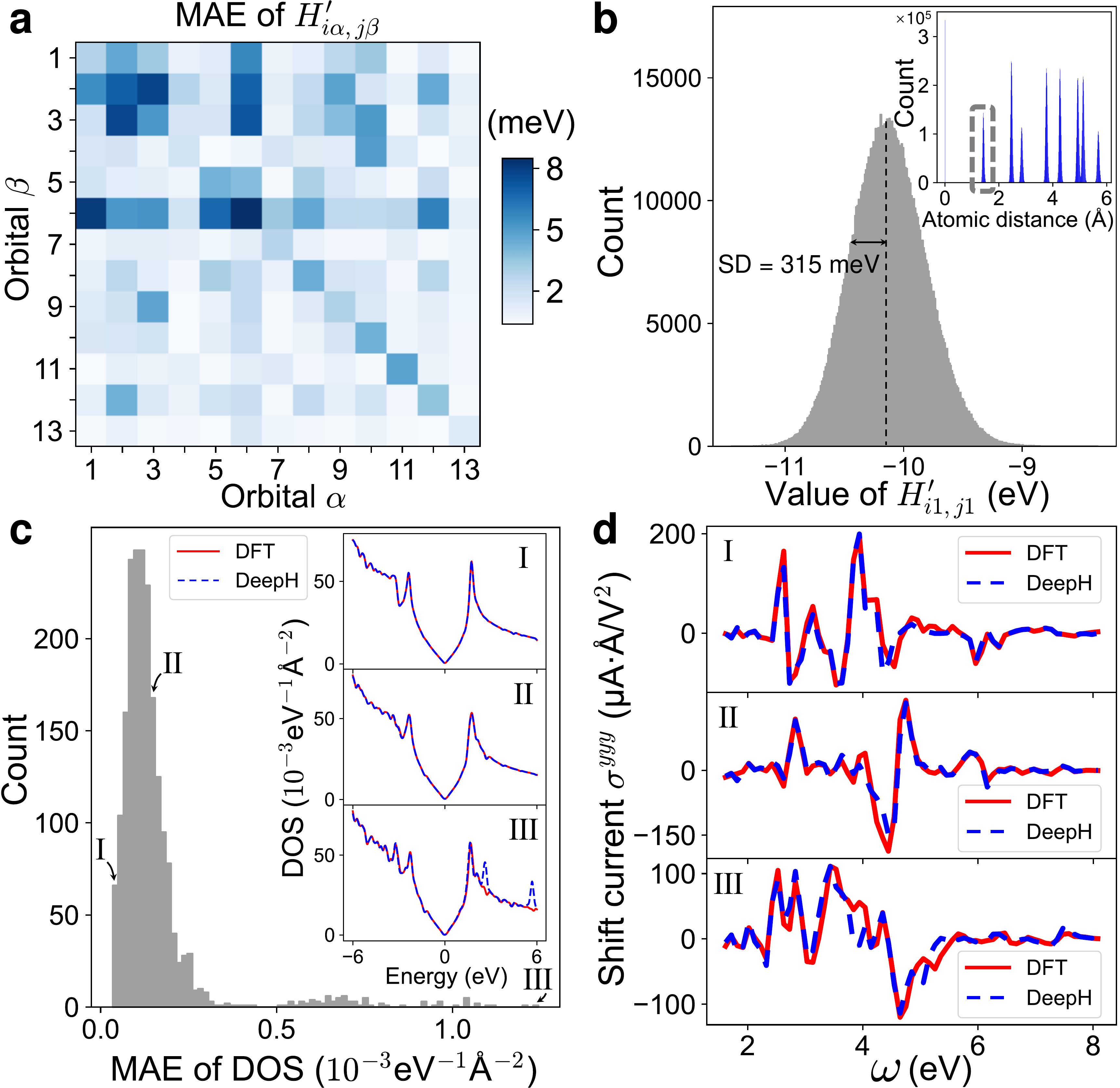}
\caption{Performance of DeepH on studying graphene. \textbf{a} MAE of $H_{i\alpha, j\beta}^\prime$ for different orbitals. \textbf{b} Distribution of $H^\prime_{i1,j1}$ for the nearest neighbors (atomic distance between 1.28 to 1.6 \AA, see the corresponding distribution in the inset). The standard deviation (SD) of computed $H^\prime_{i1,j1}$ is 315 meV for the test set. \textbf{c} Distribution of generalization MAE of DOS for 2,000 unseen material structures. Three typical structures with the best, median, and worst MAE for DOS (atomic structures included in Supplementary Data 1) are selected. Their DOS (inset) and shift current conductivity $\sigma^{yyy}$ (\textbf{d}) computed by DFT and DeepH are compared.}
\label{fig3}
\end{figure}

The training of neural network generally demands a large amount of data. In our study, 5,000 configurations of the graphene $6 \times 6$ supercell are generated by {\it ab initio} MD at a temperature of 300 K, giving 14,400,000 nonzero Hamiltonian matrix blocks. 270 configurations are used for training, which is large enough to ensure convergence as demonstrated by the calculated learning curve as a function of training set size (Supplementary Fig. 4), 90 configurations for hyperparameter optimization and the remaining for test. The mean absolute error (MAE) of $H_{i\alpha, j\beta}^\prime$ for the test set is shown in Fig. \ref{fig3}a. The MAE value averaged over all the $13 \times 13$ orbital combinations is 2.1 meV, with individual values distributed between 0.4 meV and 8.5 meV. Such a MAE is quite small considering that the Hamiltonian matrix element is typically on the order of eV. For instance, $H_{i\alpha, j\beta}^\prime$ for the $1s$ orbital and the nearest neighbor obtained from DFT calculations has a mean value of -10.1 eV and a standard deviation (SD) of 315 meV (Fig. \ref{fig3}b), whereas the corresponding MAE of DeepH is 6.6 meV, corresponding to a high coefficient of determination $r^2 = 0.9994$. For another 2,000 unseen configurations of graphene supercell sampled by {\it ab initio} MD from 100 K to 400 K, the generalization MAE of $H_{i\alpha, j\beta}^\prime$ is as small as 1.9 meV on average, demonstrating high accuracy of DeepH.

Figures \ref{fig3}c and \ref{fig3}d show results of DOS and shift current conductivity, respectively. For the 2,000 unseen configurations of graphene, the MAE between the predicted and calculated DOS with 500 points between $-6$ eV and $+6$ eV around the Fermi level is on the order of 0.1 (in a unit of $10^{-3}$eV$^{-1}$\AA$^{-2}$), much smaller than the absolute values (usually $>$10). The spectra of DOS and shift conductivity are compared between DeepH and DFT, showing satisfactory agreement.

\begin{figure}[h!]
\includegraphics[width=1.0\linewidth]{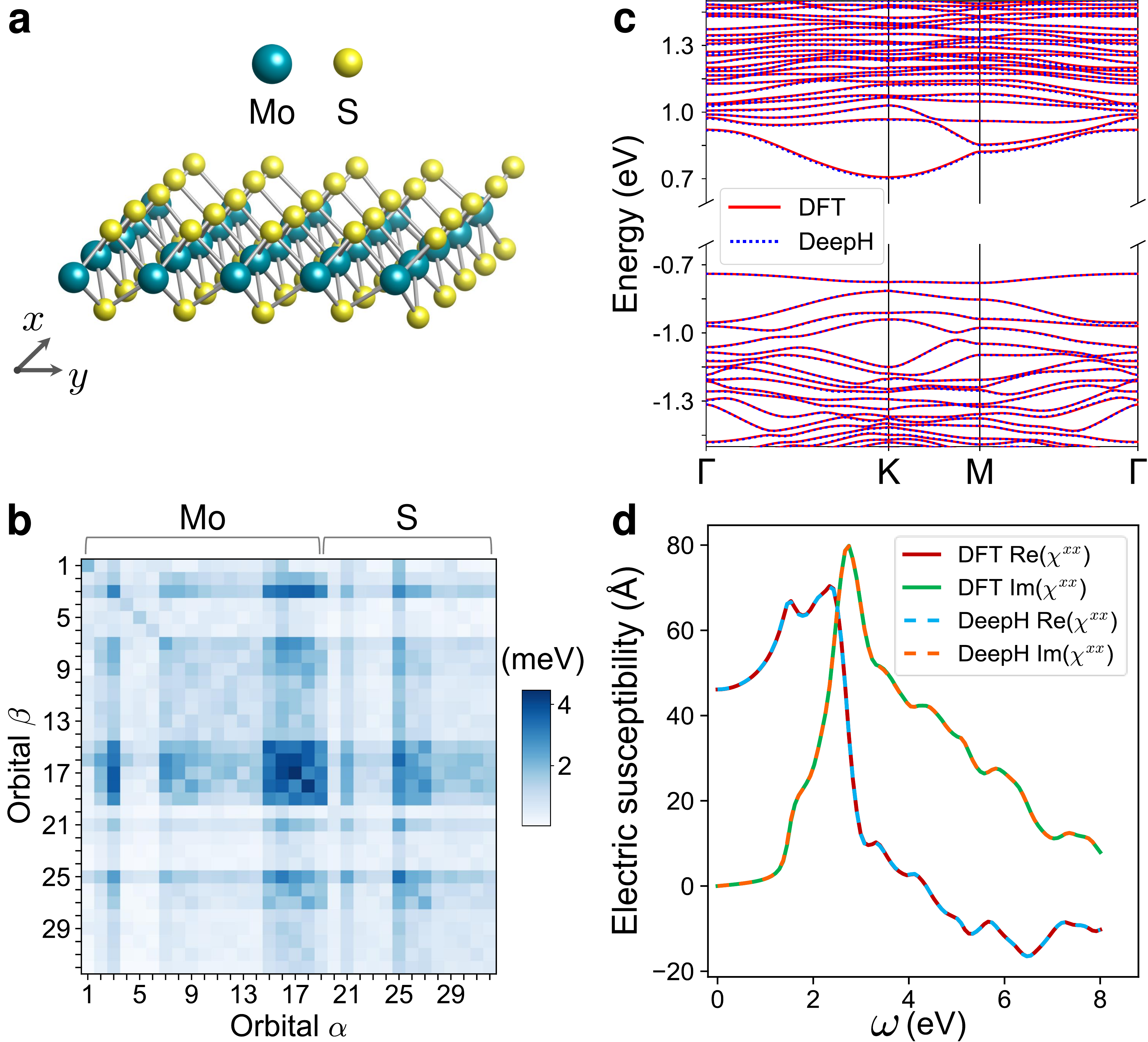}
\caption{Performance of DeepH on studying monolayer MoS$_2$. \textbf{a} Atomic structure of MoS$_2$. \textbf{b} MAE of $H_{i\alpha, j\beta}^\prime$ for different orbitals. \textbf{c} Band structures and \textbf{d} real and imaginary parts of electric susceptibility $\chi^{xx}$ computed by DFT and DeepH for a $5 \times 5$ MoS$_2$ supercell. A representative structure with median generalization MAE of DFT Hamiltonian matrix (atomic structure included in Supplementary Data 2) is considered in \textbf{c} and \textbf{d}. $\Gamma$, K, and M represent different high symmetry $k$-points of the Brillouin zone (the same applies hereinafter).}
\label{fig4}
\end{figure}

DeepH uses the embedding of atomic numbers as initial vertex features and can naturally work for systems containing multiple atomic types. For demonstration, we perform calculations on monolayer MoS$_2$, use 300 random $5\times5$ MoS$_2$ supercell structures for training, and achieve high accuracy (Fig. \ref{fig4}b-d). Specifically, the averaged MAE of $H_{i\alpha, j\beta}^\prime$ for Mo-Mo, Mo-S, S-Mo and S-S atom pairs are as low as 1.3, 1.0, 0.7 and 0.8 meV, respectively. The predicted material properties (band structure, electric susceptibility, and shift current conductivity) match well with DFT self-consistent calculations (Fig. \ref{fig4} and Supplementary Figs. 6-8). The results indicate that DeepH works well for systems containing multiple atomic types at no obvious expense of increasing computational complexity.

\begin{figure}[h!]
\includegraphics[width=0.8\linewidth]{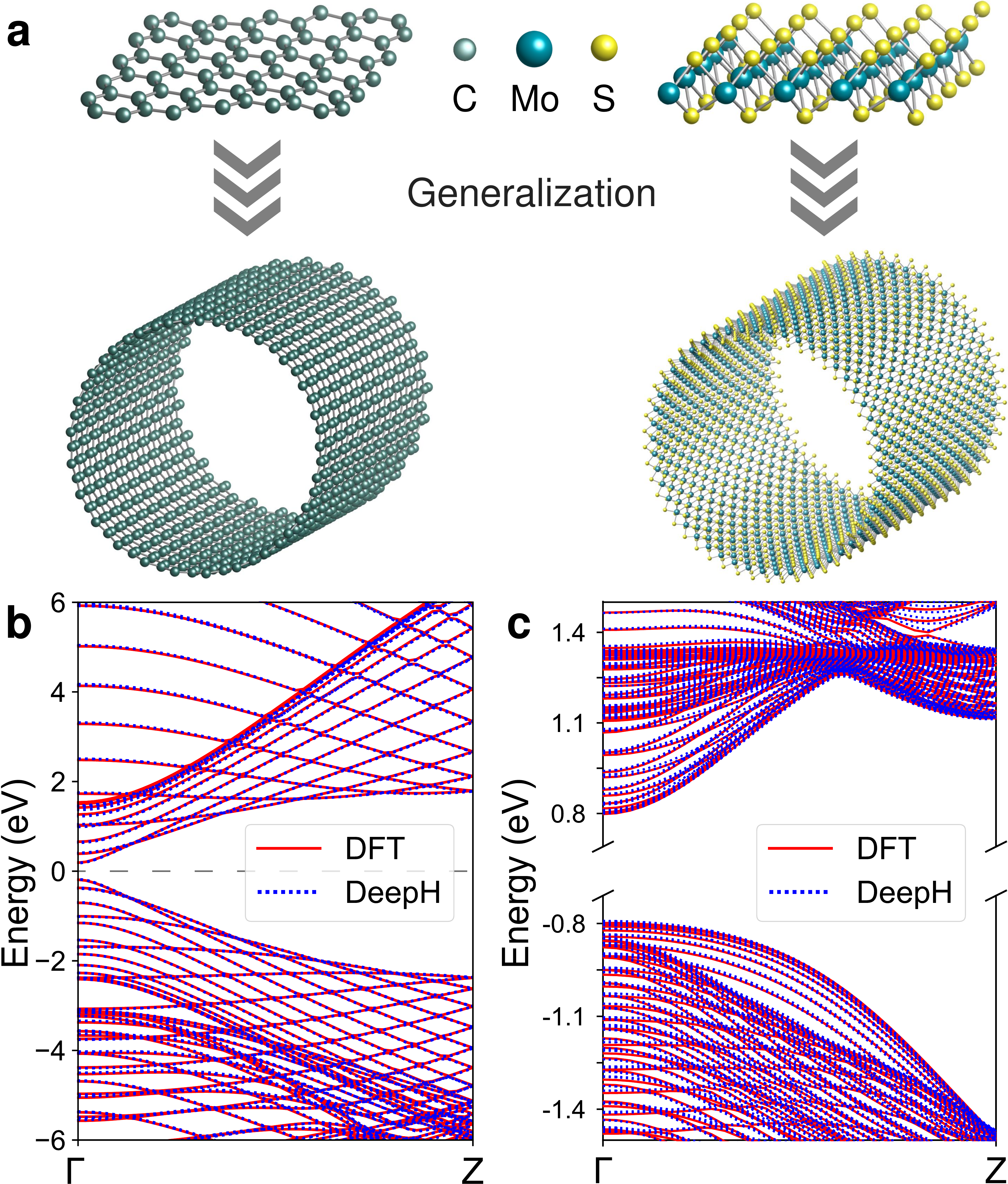}
\caption{Generalization ability of DeepH: from flat sheets to curved nanotubes. \textbf{a} Using DeepH trained by DFT results of flat sheets to study curved nanotubes for graphene (left panel) and MoS$_2$ (right panel). \textbf{b,c} Band structures for \textbf{b} a zigzag (25, 0) carbon nanotube and \textbf{c} a zigzag (50, 0) MoS$_2$ nanotube computed by DFT and DeepH. The Fermi level is aligned at the middle of the band gap.}
\label{fig5}
\end{figure}

Furthermore we test the generalization ability of DeepH by making predictions on new structures unseen in the training set (Fig. \ref{fig5}a). Testing samples of carbon nanotubes (CNTs) and MoS$_2$ nanotubes are selected for the purpose, which have curved geometry, suitable for checking the rotation covariance of the method. For CNTs, the averaged MAE of DFT Hamiltonian matrix is insensitive to nanotube chirality and monotonically decreases with increasing nanotube diameter $d$, which reduces to below 3.5 meV for $d > 2$ nm (Supplementary Fig. 5a). For a zigzag (25, 0) CNT ($d \sim 2$ nm), the predicted band structure (Fig. \ref{fig5}b) and other physical properties (like electric susceptibility Supplementary Fig. 5b) can reproduce the DFT calculation results well. Similar results are obtained for a large-diameter MoS$_2$ nanotube (Fig. \ref{fig5}c). Note that it is computationally very expensive to study large-diameter nanotubes by DFT. In contrast, their physical properties can be accurately predicted by DeepH at much lower computational expense.

Moreover, we compare the computational cost of DFT and DeepH to construct DFT Hamiltonian matrices for flat supercells and curved nanotubes of graphene and MoS$_2$ (Supplementary Table 1). Compared to DFT whose computational time roughly grows cubically with the system size, the scaling of DeepH is linear time complexity and the prefactor is much smaller. For the example study of MoS$_2$ $35\times35$ supercell, DeepH reduces computation time by three orders of magnitude. The improvement would get even more considerable for increasing system size. Therefore, the high efficiency of DeepH in dealing with large-scale material systems is demonstrated.

\subsection{Application to twisted van der Waals materials}

Twisted bilayer graphene (TBG) or twisted vdW materials in general are rising stars of materials science, where the ``magic'' Moir\'e twist provides opportunities to explore exotic quantum phases, such as correlated insulator, unconventional superconductivity, (fractional) Chern insulator, etc~\cite{Bistritzer2011, Cao2018, Cao2018_2, Yankowitz2019, Xie2021}. Despite the enormous impact, investigating the twist-angle dependence of material properties remains a grand challenge for both experiment and theory. Theoretically, empirical tight-binding and continuum models work well for simple model systems of TBG~\cite{Bistritzer2011}, but are usually not accurate enough to study other materials; {\it ab initio} calculations are demanded to accurately describe the electronic structure, but only applicable to small Moir\'e supercells. In short, the theoretical study of twisted vdW materials is limited by the accuracy-efficiency dilemma~\cite{Carr2020}. DeepH is designed to solve the dilemma, which works extremely well for studying twisted materials as we will show.

\begin{figure*}[h!]
\includegraphics[width=1\linewidth]{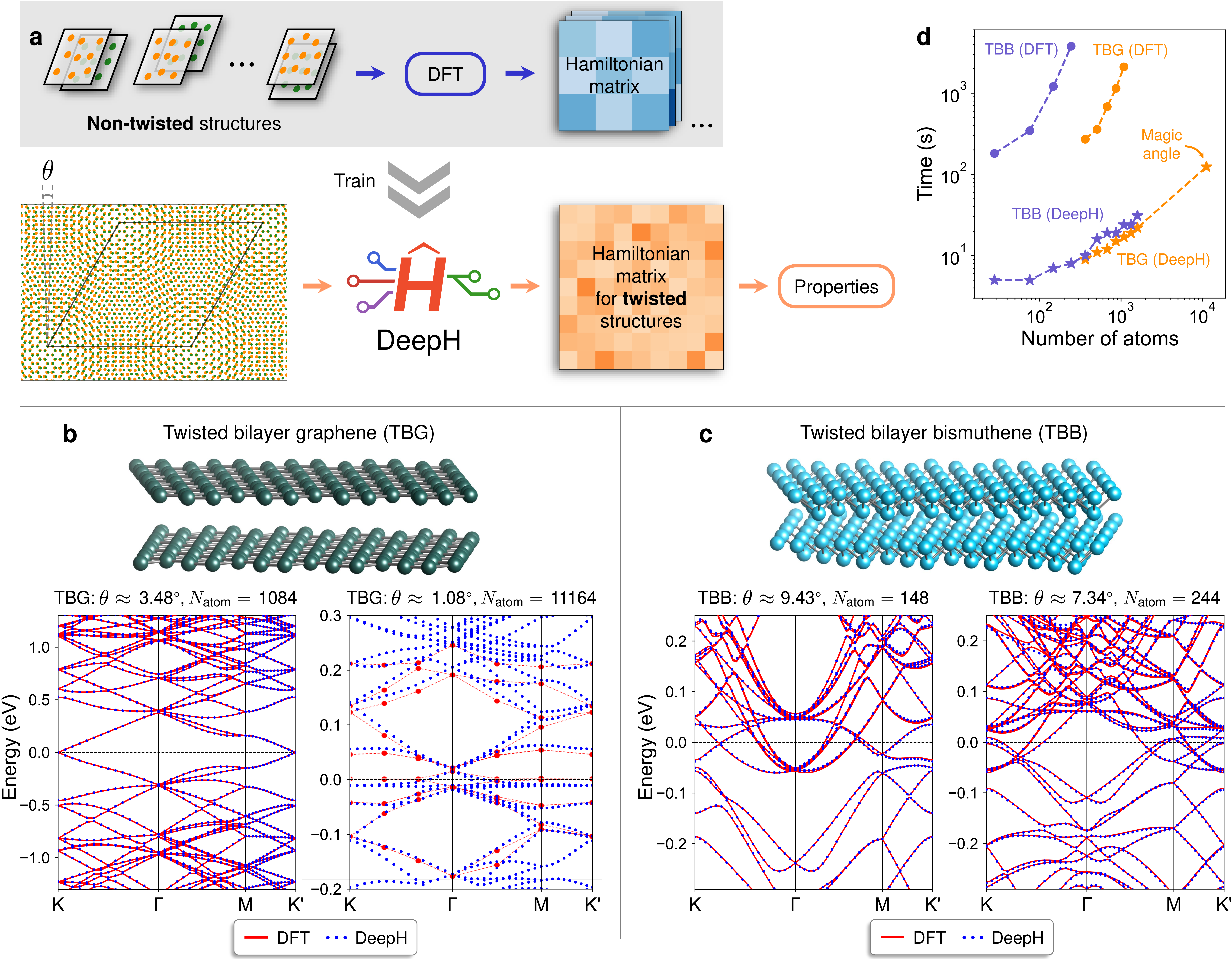}
\caption{Application of the DeepH method to study Moir\'e-twisted materials. \textbf{a} Workflow of studying twisted materials by DeepH, which uses DFT results of small non-twisted structures as training data and makes predictions on twisted structures with arbitrary twist angle $\theta$. \textbf{b,c} Band structures computed by DFT and DeepH for twisted bilayer graphenes (TBGs) (\textbf{b}: $\theta\approx 3.48^{\circ}, 1.08^{\circ}$) and twisted bilayer bismuthenes (TBBs) (\textbf{c}: $\theta \approx 9.43^{\circ}, 7.34^{\circ}$). In \textbf{c} DFT bands for magic angle $\theta\approx$ 1.08$^{\circ}$ are adapted from Ref. \cite{Lucignano2019}. \textbf{d} Computation time to construct the DFT Hamiltonian matrices of TBGs and TBBs with varying system size by DFT self-consistent calculation vs. by DeepH. For comparison the calculations are all done by one compute node equipped with 2 AMD EPYC 7542 CPUs, although DeepH works much more efficiently on GPU nodes.}
\label{fig6}
\end{figure*}

The workflow of using DeepH to study twisted materials is displayed in Fig. \ref{fig6}a. Firstly, the training data are obtained by DFT calculations of non-twisted structures, which usually contain hundreds of randomly perturbed samples of a relatively small supercell. The process of generating datasets is largely simplified because there is no need to consider varying twist angles for training. Secondly, the neural network of DeepH is trained by using the DFT data. Finally, the trained DeepH is applied to predict DFT Hamiltonian and calculate material properties for new twisted structures with an arbitrary twist angle $\theta$.  

As proof of principle, we first consider TBGs that have already been intensively studied~\cite{Bistritzer2011, Cao2018, Cao2018_2, Yankowitz2019, Xie2021}. The neural network of DeepH, once trained by DFT data for zero-twist angle, is able to give highly accurate predictions on material properties for varying twist angles. The good transferability of DeepH is demonstrated by comparing with DFT-calculated results. The averaged MAE of $H'_{i\alpha, j\beta}$ is as low as sub meV for testing Moir\'e-twisted supercells up to $\sim$1000 atoms (Supplementary Fig. 9). Due to the high accuracy in predicting DFT Hamiltonian, the calculated band structures from DeepH match well with DFT results (Fig. \ref{fig6}b and Supplementary Fig. 10) and thereby similar agreements are expected for other material properties. Noticeably, it is extremely difficult to study TBG at the magic angle $\theta\approx$ 1.08$^{\circ}$ including 11,164 atoms per supercell by traditional DFT methods, but quite easily by DeepH. For this large-size structure, the uncertainty of an ensemble of neural networks can serve as a reliability indicator of accuracy~\cite{Jeong2020}. The corresponding results indicate that the high prediction accuracy preserves for the magic-angle TBG (Supplementary Fig. 11). Indeed, the band structure calculated by DeepH matches satisfactorily with the DFT benchmark result~\cite{Lucignano2019} obtained by using the plane-wave basis at enormous computational cost (right panel of Fig. \ref{fig6}b). Importantly the existence of flat bands near the Fermi level as a characteristic feature of magic angle is well reproduced by DeepH.

Our method not only works well for TBGs, but also for other twisted vdW materials. Special attention has been paid to materials with strong SOC, such as twisted bilayer bismuthenes (TBBs), where the interplay between strong SOC and Moir\'e twist induces exotic physical properties~\cite{David2019, Gou2020}. In contrast to the Hamiltonian matrix calculated without SOC, the DFT Hamiltonian matrix with SOC has complex values and needs to take the spin degree of freedom into account for rotation transformation. Despite the additional complexity, high prediction accuracy comparable to that of TBGs is achieved for TBBs on predicting DFT Hamiltonian (Supplementary Fig. 13) as well as on calculating material properties (Fig. \ref{fig6}c and Supplementary Fig. 14).

It is worth noting that the computational time can be considerably reduced by replacing DFT self-consistent field iterations with DeepH, making {\it ab initio} electronic-structure calculation much more efficient and applicable to much larger material systems (Fig. \ref{fig6}d), such as the magic-angle TBG. On the other hand, compared to empirical tight-binding and continuum models, DeepH has slightly lower efficiency but much better accuracy and transferability. Moreover, superior to empirical methods, DeepH can easily and appropriately treat SOC, advantageous for exploring spin-related or topological quantum phenomena. For comparison, the performance of different theoretical methods is summarized in Supplementary Table 5. DeepH can outperform the currently used approaches in studying twisted materials. The method is promising for studying twist-angle dependent physical properties and for building twisted materials databases.

\subsection{Wide applicability of DeepH}

Up to now many kinds of deep-learning DFT methods have been developed~\cite{Behler2007, Schutt2018, Zhang2018, Klicpera2020, Brockherde2017, Grisafi2019, Chandrasekaran2019, Tsubaki2020, Schutt2019, Unke2021, Grisafi2018, Gu2020, Nagai2020, Dick2020, Kirkpatrick2021, Mills2019, Zubatiuk2021}. They can be classified into two groups, aiming to improve either accuracy or efficiency of DFT by deep learning techniques. As representative works of the former group, substantial breakthroughs have been achieved recently in developing advanced exchange and correlation functionals via deep neural network~\cite{Nagai2020, Dick2020, Kirkpatrick2021}. The latter group of works try to reproduce DFT results via deep learning, similar as DeepH. Among them, great successes have been achieved on deep-learning potential~\cite{Behler2007, Schutt2018, Zhang2018, Klicpera2020}, which facilitate highly efficient {\it ab initio} atomic-structure calculations. However, the corresponding developments of {\it ab initio} electronic-structure methods are preliminary. Most current works select a one-step strategy and directly learned individual physical quantities, such as band gap, band dispersion, electron density, and wavefunction~\cite{Brockherde2017, Grisafi2019, Chandrasekaran2019, Tsubaki2020, Grisafi2018, Gu2020}. For comparison, a two-step strategy employed by DeepH, which first learns the DFT Hamiltonian and then predicts the desired physical properties, is more advantageous in two aspects. Firstly, all the above mentioned electron-related physical quantities can be simultaneously derived from DeepH. Secondly and more importantly, the complex structure-property relation can be accurately described by DeepH as we demonstrated here, benefiting from the nearsighted nature of DFT Hamiltonian. In contrast, the nearsightedness principle is not applicable to some physical quantities, such as band structure and wavefunction.

Recently, Hegde and Bowen attempted to study the DFT Hamiltonian by statistical learning (not deep learning)~\cite{Hegde2017}, which works for small unit cells of simple metal copper. However, this method can hardly be applied to study more complex material systems due to the limited expressive power of statistical learning and the lack of an appropriate treatment of rotation covariance. A quantitative comparison between this method and DeepH is presented in Supplementary Section 6~\cite{supp}. A few primary deep-learning works have done for small molecules~\cite{Schutt2019, Unke2021}, which are applicable to systems with fixed number of atoms. Distinguished from existing deep-learning DFT methods, the DeepH method we developed shows excellent performance on studying periodic or non-periodic crystalline materials in terms of accuracy, efficiency and transferability as demonstrated by case studies on various kinds of quasi-one-dimensional (quasi-1D) and two-dimensional (2D) materials without or with multiple elements, curved geometry, or Moir\'e twist. DeepH can also be applied to study material systems of other space dimensions. For instance, we have made experiments on 3D bulk materials (including silicon and allotropes of carbon) as well as quasi-0D molecules (Supplementary Section 6~\cite{supp}). With the help of DeepH, the accuracy-efficiency dilemma of DFT can be solved and efficient {\it ab initio} electronic structure calculations become feasible for large-scale material systems.

One may straightforwardly check the generalization ability of DeepH by performing principal component analysis (PCA) for the output atom features of the final MP layer or the output bond features of the final LCMP layer. We performed PCA on monolayer sheets versus nanotubes and non-twisted versus twisted bilayers, as presented in Supplementary Figs. 17-20. The corresponding results are discussed in Supplementary Section 5~\cite{supp}. It is learned from PCA that DeepH can make accurate predictions on new structures with principal components significantly different from the original training set, showing satisfactory generalization ability.

It is worth while comparing our method with covariant neural network methods (such as Tensor-field networks~\cite{Nathaniel2018}, Cormorant~\cite{Anderson2019}, PhiSNet~\cite{Unke2021}, etc.), which are based on spherical harmonic functions and group representation theory. These methods require tensor products using Clebsch-Gordon coefficients in every layer of neural network during training and inference processes to ensure rotational covariance. The tensor-product computation could be very expensive, especially for large-size systems and for calculations involving basis sets of high orbital angular momenta. As far as we know, applying these methods to study electronic structure of large-scale material systems remains elusive.

In contrast, our method only needs to perform the basis transformation once before training process, which is computationally very efficient. Moreover, benefitting from the rotation invariant nature of local coordinates, our approach can apply rotation invariant neural network to predict rotation covariant quantities, making the neural network architecture more flexible and efficient. Importantly, further development of the method would benefit from the great developments of transformation invariant neural network. Since all the important local bonding information, including bond length and orientation information, have been included as input, our method is expressive enough to achieve high prediction accuracy. Quantitative comparisons against the tensor-product-based method on studying molecule datasets~\cite{Schutt2019, Unke2021} indicate that DeepH can achieve comparable accuracy with much less computation time and number of parameters (Supplementary Table 4).

Deep neural network in principle can be applied to deal with complex problems with large configuration space due to its remarkable expressive power. The object of the present work is to learn the DFT Hamiltonian matrix as a function of atomic positions. For most physical problems, only atomic configurations near equilibrium positions are concerned thermodynamically due to their relatively low energies. Thus we focused on configuration space near equilibrium for a given material. Solids with nearly periodic structures (like graphene with lattice vibrations) usually have small configuration space. The DeepH method can work well for such kind of systems. On the other hand, training a model in a large configuration space is indeed much more challenging, which usually requires more training data and possibly demands methodological improvement to achieve good accuracy.

To test the performance of DeepH, we considered two kinds of material systems with a relatively larger configuration space: (i) 3D bulk structures including different allotropes of carbon (i.e., graphite and diamond) and (ii) quasi-0D molecules. For the former, one unified neural network is applied to predict DFT Hamiltonian for two kinds of carbon allotropes. The MAEs of DeepH do not increase with respect to that of graphene, though the configuration space gets larger. For the latter, we studied molecules of increasingly larger size (from 3 to 21 atoms) for considering the growth of configuration space. Their averaged MAEs of $H'_{i\alpha, j\beta}$ are on the order of sub meV, also lower than that for graphene. More detailed results are presented in Supplementary Section 4 and 6~\cite{supp}. These experiments suggest that DeepH is very likely applicable to the study of material systems spanning a large configuration space. We would like to do more critical experiments and developments in future works.

\section{Discussion}
We have proposed a general framework to represent DFT Hamiltonian by deep neural network, which builds universally a mapping from materials structures to physical properties. The method extends the scope of first-principles research and opens opportunities to investigate fundamental physics and large-scale material systems, such as twisted vdW materials. However, the current DeepH framework is not without limitations. For instance, the trained model is only applied to study unseen materials that have close chemical bonding environment with dataset. To investigate material systems with strongly varying chemical environment, one still needs to manually design an appropriate dataset for improving training efficiency. An automatic construction of dataset and on-the-fly optimization of training process could be explored in the future.

Some generalizations of the method are straightforward, whereas some others are not. For instance, the method can be generalized to study large-scale systems without periodicity (e.g., non-commensurate twisted materials, quasi-crystals). Some other material systems (e.g., disorder, defect~\cite{Mills2019}, interface) in principle can be described by DeepH as well, but demand more training data to learn varying chemical environment. Moreover, DeepH is compatible with DFT not only for exchange correlation functionals in the local density approximation or generalized gradient approximation (GGA) but also for the more advanced functionals, such as meta-GGA, hybrid functionals, etc. Note that the hybrid functionals demand a larger cutoff radius for constructing crystal graphs than usual. Furthermore, by combining deep-learning potential and DeepH together, efficient MD simulation and electronic-structure calculation can be performed simultaneously, making the real-time simulation of electron-lattice coupling possible. Another valuable extension of this current work is the combination of DeepH and efficient linear algebra algorithms (e.g., diagonalization for large sparse matrices, linear algebra algorithms on GPUs), which could further improve computational efficiency and promote the exploration of mesoscopic physics and materials. There exist much room for future development of the method, which we would like to consider in following works.

\section{Methods}

\subsection{Dataset preparation}

We generate random structural configurations of 6 $\times$ 6 monolayer graphene and 5 $\times$ 5 monolayer MoS$_2$ supercells by {\it ab initio} MD calculations via Vienna {\it ab initio} simulation package~\cite{Kresse1996}. Simulations are performed with the projector-augmented wave~\cite{Blochl1994,Kresse1999}  pseudopotentials and the GGA parameterized by Perdew, Berke and Ernzerhof (PBE)~\cite{Perdew1996}. The cut-off energy of plane waves is 450 eV and only the $\Gamma$ point is used in our $k$-mesh. For monolayer graphene supercells, two simulations are carried out under the canonical ensemble: one with a constant temperature of 300 K, and the other with temperature increasing from 100 K to 400 K. With the time step of 1\,fs, our dataset consists of 5,000 frames obtained at 300 K and 2,000 frames obtained between 100 K and 400 K. As for monolayer MoS$_2$ supercells, 1,000 random atomic structures of $5 \times 5$ supercells are generated by {\it ab initio} MD calculations performed at 300 K with the time step of 1\,fs.

Furthermore, in order to train DeepH models for Moir\'e-twisted vdW materials, we prepare the datasets for TBGs (TBBs) from zero-twist-angle 4 $\times$ 4 (3 $\times$ 3) bilayer supercells by shifting one of the two vdW layers within the 2D plane and subsequently inserting random perturbations to each atomic sites. The interlayer spacing of the fully relaxed bilayer unit cells with the most energetically favorable stacking is used to construct the training dataset and Moir\'e-twisted supercells (3.35 \AA\ for TBG and 3.20 \AA\ for TBB). In total, 300 and 576 shifted and perturbed supercell structures are included in datasets for TBG and TBB, respectively.

We calculate DFT Hamiltonians with pseudo-atomic localized basis functions as implemented in OpenMX software package version 3.9~\cite{Ozaki2003, Ozaki2004}. Calculations are performed with PBE exchange correlation functional and norm-conserving pseudopotentials~\cite{Morrison1993}. For monolayer graphene, CNTs, and TBGs, C6.0-$s$2$p$2$d$1 pseudo-atomic orbitals are used, including 13 atomic-like basis functions, with the cut-off radius $R_C =$ 6.0 Bohr. For monolayer MoS$_2$ and MoS$_2$ nanotubes, Mo7.0-$s$3$p$2$d$2 and S7.0-$s$2$p$2$d$1 pseudo-atomic orbitals are used, including 19 atomic-like basis functions for molybdenum and 13 for sulfur ($R_C =$ 7.0 Bohr). For TBBs, Bi8.0-$s$3$p$2$d$2 pseudo-atomic orbitals are used, including 19 atomic-like basis functions ($R_C =$ 8.0 Bohr). The energy cutoff is set to 300 Ry. A Monkhorst-Pack $k$-mesh of 5 $\times$ 5 $\times$ 1 is used for supercells of monolayer graphene with 72 atoms, monolayer MoS$_2$ with 75 atoms, bilayer graphene with 64 atoms, and bilayer bismuthene with 36 atoms. For supercells with atom number larger than 1,000, only $\Gamma$ point was used. Meanwhile, a Monkhorst-Pack $k$-mesh of 1 $\times$ 1 $\times$ 29 is used for CNTs and MoS$_2$ nanotubes, then 1 $\times$ 1 $\times$ 1 (2 $\times$ 2 $\times$ 1) for TBGs (TBBs). The SOC is considered in the calculation of bilayer bismuthene supercells and TBBs.

\subsection{Physical properties derived from DFT Hamiltonian}

In a non-orthogonal atomic orbital basis set, the Hamiltonian and overlap matrix elements are defined as
\begin{align}
H_{i\alpha, j\beta} = \langle\phi_{i\alpha}|\hat{H}|\phi_{j\beta}\rangle
\end{align}
and
\begin{align}
S_{i\alpha, j\beta} = \langle\phi_{i\alpha}|\phi_{j\beta}\rangle ,
\end{align}
where $|\phi_{i\alpha}\rangle$ denotes the atomic orbital $\alpha$ of atom $i$. 
DFT Hamiltonian matrix can be obtained from DFT self-consistent field calculations or predicted by DeepH method. The overlap matrix is obtained by the inner product of the basis at very low computational cost. Thus it is unnecessary to learn this quantity by neural network. After Fourier transformations of Hamiltonian and overlap matrices, the eigenvalues $\mathcal{E}_{n\mathbf{k}}$ and the eigenstates $v_{n\mathbf{k}}$ of the Hamiltonian $\hat{H}$ at band $n$ and wavevector $\mathbf{k}$ can be obtained by solving the generalized eigenvalue problem~\cite{Wang2019}
\begin{align}
H(\mathbf{k})v_{n\mathbf{k}} = \mathcal{E}_{n\mathbf{k}}S(\mathbf{k})v_{n\mathbf{k}} .
\end{align}
For Moir\'e-twisted materials in the current study, ARPACK library is used to compute a few eigenvalues of the large-scale sparse Hamiltonian matrix obtained from the DeepH method.

The 3D electric susceptibility $\chi$ and shift current conductivity $\sigma$ \cite{Sipe2000} as functions of light frequency $\omega$ are given by
\begin{align}
\label{chi}
\chi^{ab} = \frac{e^2}{\epsilon_0\hbar}\int\frac{d^3\mathbf{k}}{(2\pi)^3}\sum_{n, m}f_{nm}\frac{r^a_{nm}r^b_{mn}}{\omega_{mn}(\mathbf{k}) - \omega - i \eta}
\end{align}
and
\begin{align}
\label{sigma}
\sigma^{abc}(\omega) = &\frac{\pi e^3}{\hbar^2}\int\frac{d^3\mathbf{k}}{(2\pi)^3}\\
&\times\sum_{n, m} f_{nm}
\mathrm{Im}\left(r^{b}_{mn}r^{c;a}_{nm} + 
r^{c}_{mn}r^{b;a}_{nm}\right)
\delta\left(\omega_{mn}(\mathbf{k}) - \omega\right) ,\nonumber
\end{align}
where $a,b,c$ are Cartesian directions, $\epsilon_0$ is the vacuum permittivity, $\hbar$ is the reduced Planck's constant, $e$ is the charge of electron, and $\eta$ is an infinitesimal relaxation rate. $\omega_{nm}(\mathbf{k}) = \frac{E_{n\mathbf{k}} - E_{m\mathbf{k}}}{\hbar}$ and $f_{nm} = f_n(\mathbf{k}) - f_m(\mathbf{k})$ are the difference of energy eigenvalues and Fermi-Dirac occupations of bands $n$ and $m$ at wave vector $\mathbf{k}$, respectively. $r^a_{nm}$ and $r^{b;a}_{nm} = \frac{\partial r^b_{nm}}{\partial k^a} - i(r^a_{nn} - r^a_{mm})r^b_{nm}$ are Berry connection and its general derivative, which are calculated with DFT Hamiltonian using the method developed in Ref. \cite{Wang2019}.

For low dimensional systems, the response functions calculated by Eqs. \eqref{chi} and \eqref{sigma} need to be redefined to exclude the influence of vacuum layer in the supercell. As we are interested in the susceptibility of 2D MoS$_2$ layers and quasi-1D CNTs and the shift current conductivity of 2D graphene layers, the 2D susceptibility, 1D susceptibility, and the 2D sheet conductivity are given by
\begin{align}
\chi_{\text{2D}} &= L_{\text{sp}} \cdot \chi_{\text{3D}} ,\\
\chi_{\text{1D}}^{\mathrel{/\mskip-4mu/}} &= S_{\text{sp}} \cdot \chi^{\mathrel{/\mskip-4mu/}}_{\text{3D}} ,\\
\end{align}
and
\begin{align}
\sigma_{\text{2D}} &= L_{\text{sp}} \cdot \sigma_{\text{3D}},
\end{align}
respectively, where $S_{\text{sp}}$ and $L_{\text{sp}}$ are the cross section area and height of the supercell.  $\chi^{\mathrel{/\mskip-4mu/}}_{\text{3d}}$ is the electric susceptibility along the periodic direction.

\subsection{Details on training neural network}

Eqs. \eqref{vertex_update} and \eqref{edge_update} include neural networks for updating vertex feature and edge feature. The neural network of vertex $\Phi^{(l)}_v(x) = \sigma(xW^{(l)}_1 + b^{(l)}_1)\odot g(xW^{(l)}_2 + b^{(l)}_2)$, where the input $x\in\mathbb{R}^{n_\text{in}}$, the weight $W\in\mathbb{R}^{n_{\text{in}}\times n_{\text{in}}}$, the bias $b\in \mathbb{R}^{n_\text{out}}$, $\odot$ denotes element-wise multiplication, $\sigma$ denotes the sigmoid function, and $g$ denotes the softplus function~\cite{Xie2018}. The neural network of edge $\Phi^{(l)}_e(x) = \text{SiLU}\left(xW^{(l)}_3 + b^{(l)}_3\right) W^{(l)}_4 + b^{(l)}_4$, which is a fully connected neural network including a hidden layer and a SiLU activation function.

The MPNN model we use includes five MP layers, one LCMP layer and thus $471409 + 129 \times N_{\text{out}}$ parameters, where $N_{\text{out}}$ is a hyperparameter of the number of selected orbital pairs. The cutoff radius $R_C$ for constructing crystal graphs is set to the cutoff radius of corresponding atomic-like orbitals. The dimension of elemental embeddings, as well as vertex feature vectors in each layer is set to 64. The initial edge features are a set of 128 Gaussian functions $\exp(-(\left|r_{ij}\right| - r_n)^2/\sigma^2)$ where the center $r_n$ is placed linearly between 0 and 6 \AA, and $\sigma^2$ is set to 0.0044. The edge feature vector in each layer is a 128-dimensional vector. There are 25 real spherical harmonic functions $\{Y_{Jm}\}$ to expand orientation information, where $J$ is an integer ranging from 0 to 4, $m$ is an integer between $-J$ and $J$. The batch size of 12, 3, 4 and 1 is set for monolayer graphene, monolayer MoS$_2$, TBG and TBB, respectively. An Adam optimizer is used with a learning rate initiated at $1\times 10^{-3}$ which later reduces to $2\times 10^{-4}$ and finally to $4\times 10^{-5}$. We implemente the MPNN model in DeepH method using Pytorch-Geometric~\cite{Fey2019} Python library.

It is optional to learn $H_{i\alpha, j\beta}^\prime$ separately or to treat $H_{ij}^\prime$ as a whole. In the example study on monolayer graphene and TBG, multiple MPNN models are trained to represent the mapping from $\{\mathcal{R}\}_N$ to $H_{ij}^\prime$ for different orbital pairs. For the MoS$_2$ and TBB, multi-dimensional vector outputs of a single MPNN model are used to represent Hamiltonian matrix blocks as a whole to achieve high efficiency.

\section{Data availability}

Source data for Figures 3-6 is available with this paper. The dataset used to train the deep learning model is available at Zenodo~\cite{dataset}.

\section{Code availability}

The code used in the current study is available at GitHub (https://github.com/mzjb/DeepH-pack) and Zenodo~\cite{code}.

\section{Acknowledgments}

This work was supported by the Basic Science Center Project of NSFC (Grant No. 51788104), the National Science Fund for Distinguished Young Scholars (Grant No. 12025405), the National Natural Science Foundation of China (Grant No. 11874035), the Ministry of Science and Technology of China (Grants No. 2018YFA0307100 and 2018YFA0305603), the Beijing Advanced Innovation Center for Future Chip (ICFC), and the Beijing Advanced Innovation Center for Materials Genome Engineering. M.Y. was supported by Shuimu Tsinghua Scholar Program and Postdoctoral International Exchange Program. R.X. was funded by China Postdoctoral Science Foundation (Grant No. 2021TQ0187).

\section{Author contributions}

Y.X. and W.D. proposed the project and supervised H.L., Z.W., and N.Z. carrying out the research with the help of M.Y., R.X, and X.G. All authors discussed the results. Y.X. and H.L. prepared the manuscript with input from other co-authors.

\section{Competing interests}

The authors declare no competing interests.

\end{document}